\documentclass[]{elsarticle}

\usepackage{lineno,hyperref}
\usepackage{amsmath}
\modulolinenumbers[5]
\usepackage{subfigure}
\usepackage{xcolor}
\usepackage{placeins}
\usepackage{caption}

\journal{Journal}

\graphicspath{{figs/}}

\begin{document}

\begin{frontmatter}

\title{Light induced spiking of proteinoids}


\author{Panagiotis Mougkogiannis*}
\author{Andrew Adamatzky}
\address{Unconventional Computing Laboratory, UWE, Bristol, UK}

\cortext[cor1]{Corresponding author: Panagiotis.Mougkogiannis@uwe.ac.uk (Panagiotis Mougkogiannis)}

\begin{abstract}
Proteinoids, or thermal proteins, are produced by heating amino acids to their melting point and initiation of polymerisation to produce polymeric chains. In aqueous solutions proteinoids swell into hollow microspheres. These microspheres produce endogenous burst of electrical potential spikes and change patterns of their electrical activity in response to illumination. We report results of detailed investigation on the effects of white cold light on the spiking of proteinoids. We study how different types and intensities of light determine proteinoids' spiking amplitude, period, and pattern. The results of this study will be utilised to evaluate proteinoids for their potential as optical sensors and their application in unconventional computing. 
\end{abstract}


\end{frontmatter}


\section{Introduction}

Thermal proteins --- proteinoids~\cite{fox1992thermal} --- are produced by heating amino acids to their melting point, 80–200~\textsuperscript{o}C. The heating initiates the polymerisation to produce polymeric chains~\cite{harada1958thermal,fox1992thermal}.  In aqueous solutions proteinoids swell into hollow microspheres~\cite{fox1992thermal}. 
As has been reported by Przybylski in 1980s the proteinoid microspheres maintain a steady state membrane potential 20~mV to 70~mV without any stimulating current~\cite{przybylski1984physical,przybylski1985excitable}. More the microspheres exhibit action-potential like spikes. The electrical activity of the microspheres includes spontaneous bursts of electrical potential~\cite{ishima1981electrical}.  Membrane, action, and oscillatory potentials recorded from the microspheres composed of thermal protein, glycerol, and lecithin~\cite{ishima1981electrical,przybylski1982membrane} are observed for several days~\cite{bi1994evidence}. Exact mechanisms of these electrical oscillations remain unknown, yet there are theoretical models related to abrupt changes of the strength of coupling between basic and acidic proteinoids~\cite{matsuno1984electrical}, ionic gradients across the microspheres' membranes~\cite{kimizuka1964ion}  and  immobilisation of mobile ions on the membranes~\cite{tamagawa2015membrane}. 

In \cite{adamatzky2021towards} we proposed a theoretical frameworks for developing computing devices with proteinoid microspheres by considering ensembles of the microspheres as networks of coupled oscillators for mining Boolean circuits. Most efficient way for a parallel input of data into proteinoid-based computing devices would be to use light. Photosensitivity of proteinoids was firstly reported in early 1980s~\cite{ishima1981electrical,przybylski1983towards}: light initiated spiking activity which slowly extinguished after the illumination was switched off. An interesting observation, mentioned in~\cite{ishima1981electrical}, was that the  oscillations stop on the higher level of the  potential. The goal of present paper is to investigate ``lighting driven spiking of proteinoids'', a hypothesis that could significantly alter our understanding of and engagement with proteinoids. This study could contribute in the development of novel approaches to  regulate proteinoids' formation, stability, and function~\cite{nakashima2005metabolism,fox1980metabolic}. Understanding the impact of light energy in proteinoid folding could help in the development of more efficient energy harvesting and storage systems~\cite{rajput2022energy,9347297},\cite{englander2014nature}.

Research into the ability of light to influence protein interactions in living cells has gained traction in recent years. Blue light, the cryptochrome 2 and CIB1 proteins, and arabidopsis thaliana were used by C-H. Wu and colleagues to create a method for inducing protein interactions in living cells in 2010~\cite{kennedy2010rapid}. A light-controlled protein delivery method into eukaryotic cells was developed in subsequent investigations by S. Scherzinger et al. in 2020~\cite{lindner2020litesec}. Using light-switchable nanobodies, J.-P. Schatzle et al. established optogenetic regulation of protein binding in 2020~\cite{gil2020optogenetic}. This work has led to substantial progress in our knowledge of the mechanisms that regulate cellular behaviour by allowing for the exact modulation of protein interactions in living cells using light. 
The first photo aged proteolysis chimaeras (pc-PROTACs) were developed by Xue et al.~\cite{xue2019light}, and they are able to selectively cause the breakdown of target proteins in a light-dependent, reversible, and tunable manner. Since then, others have expanded upon this work to create their own light-driven proteinoid-based systems~\cite{liu2020functional}. This study laid the groundwork for future advances in light-based technologies~\cite{bao2022light} that regulate protein function.




There is great potential for progress in our understanding of biochemical oscillations, and the study of light-induced spiking of proteinoids is a relatively new area of inquiry. In this study, the effects of various lighting intensities on the behaviour of proteinoids, which are small molecules with protein-like structures, are investigated. Oscillatory behaviour, or spiking in response to light, is one of the most exciting results made on proteinoids under illumination. This pulsating behaviour changes across three unique light intensity regimes. Proteinoids show oscillatory behaviour at light intensity below a critical value. Oscillations cease when the light intensity exceeds the critical value. This points to the presence of a logical XOR gate that can toggle between two states based on the amount of light present. This research has far-reaching implications, and it shows that studying how proteinoids spike in response to light could lead to innovative advances in biotechnology and unconventional computing.

\section{Methods}

Preparation of proteinoids followed published protocols~\cite{mougkogiannis2023transfer}. PicoLog ADC-24 high resolution data logger was used to measure electrical potential oscillations (PicoLog Pico Technology, UK). 
Disposable Platinum-Iridium subdermal needle electrodes, length 12~mm, diameter 0.40~mm (CNSAC MedShop GmbH, Randersacker, Germany) inserted into the proteinoid solution were interfaced with the data logger. 

\begin{table}[!tbp]
\centering
\caption{The relative ranges of light intensity for various setting of 5~mm light guide knob of the Photonic PL 2000 source.
}
\begin{tabular}{|c|c|c|}
	\hline
Relative value & Max Illuminance (Klux)       \\
	\hline\hline
	 1 &    37.2 \\
 \hline
	 2 &    55.9 \\
 \hline
	 3 &    101.8  \\
 \hline
	 4 &   134.2  \\
 \hline
	 5 &   160.8 \\
 \hline
	 6 &   186.6 \\
	
	\hline
	\end{tabular}

\label{vhjfvahjkvf}
\end{table}

With its halogen input and 18~Mlux light intensity, the Photonic PL 2000 served as a constant cold white light source for stimulating proteinoid solutions with light.  The PL 2000 was calibrated to emit white light with a lamp voltage of 13~V. The proteinoid sample was held at a distance of 5~cm from the PL 2000.
A 3.5 $mm$ light guide can produce a maximum of 0.8 $W/cm^{2}$ of white light, whereas a 5 mm light guide can produce 0.6 $W/cm^{2}$ of white light. White light relative intensity level 1 relates to the lowest light intensity that the electronic device is capable of producing. The light intensity in relative values 1, 2, 3, 4, 5, and 6 under mechanical control are not set values, but ranges. The mechanical control is a knob that can be turned to continually alter the light intensity within each range. The relative ranges of light intensity for various light guides are listed in the Tab.~\ref{vhjfvahjkvf} below~\cite{PHOTONIC}.

\section{Results}
\subsection{Uncovering the Nanostructure of Proteinoids Through SEM Imaging}
\begin{figure}[!tbp]
\centering
\includegraphics[width=1\textwidth]{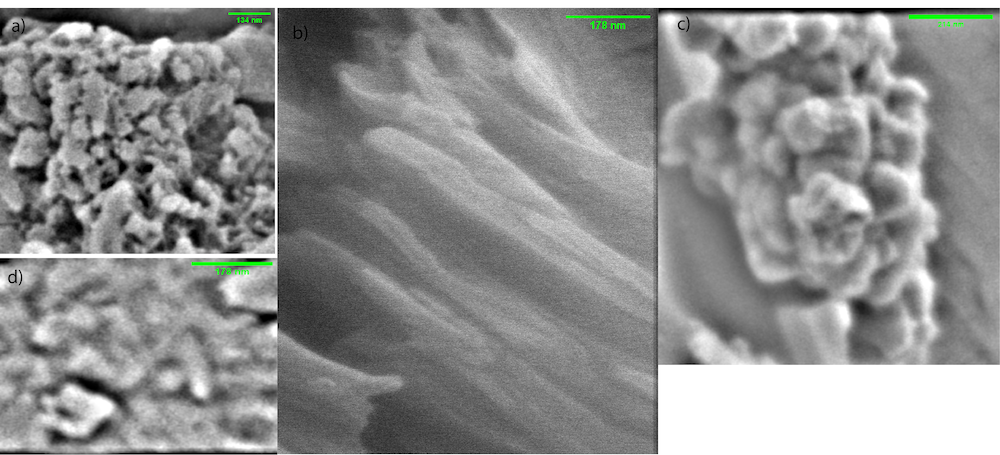}
\caption{Exploring four different proteinoids using SEM: a) 134 nm scale bar for L-Glu:L-Phe; b) 178 nm scale bar for L-Phe:L-Lys; c) 214 nm scale bar for L-Phe; and d) 214 nm scale bar for L-Glu:L-Phe:L-His-a fascinating exploration of the intricate architectures of thermal proteins. }
\label{hbjkbvjbsjVB}
\end{figure}

The results of imaging proteinoids with the FEI Quanta 650 field emission scanning electron microscope were fairly outstanding (Fig.~\ref{hbjkbvjbsjVB}). The gold-coated samples were scanned in three dimensions, with the L-Glu:L-Phe (Fig.~\ref{hbjkbvjbsjVB}a), L-Glu:L-Phe:L-His (Fig.~\ref{hbjkbvjbsjVB}b), and L-Phe (Fig.~\ref{hbjkbvjbsjVB}d) pictures displaying structures resembling nanospheres and the L-Phe:L-Lys (Fig.~\ref{hbjkbvjbsjVB}c) image displaying a neuron-like structure. 

The L-Phe:L-Lys nano-needles seen in the SEM pictures had a diameter of 33.7~nm and a length of 850.7~nm. The versatility of this structure makes it a crucial find for the advancement of nanotechnology. These nano-needles are small enough to be employed as transistor components, opening the door to their potential usage in the fabrication of nano-scale electronic devices. This has the potential to result in more efficient and less expensive fabrication methods and smaller, more powerful electronic devices. Small enough to pass through cell membranes unharmed, nano-needles of this size have potential in medical devices such drug delivery systems.

Proteinoids L-Glu:L-Phe, L-Glu:L-Phe:L-His, and L-Phe were measured to have diameters of 57.8~nm, 49.5~nm, and 95.1~nm, respectively, using scanning electron microscopy.  Diameter differences between the proteinoids L-Glu:L-Phe, L-Glu:L-Phe:L-His, and L-Phe nanospheres are indicative of their distinct functional and structural profiles. The L-Glu:L-Phe:L-His nanospheres, for instance, measured in at a smaller 49.5~nm diameter than the other two, suggesting that their denser structure makes them suitable for uses that call for a smaller footprint or distinct physical features.

The nanospheres provided a fair representation of the structure of the proteinoid, but the neuron-like structure was very intriguing. This structure indicated a potentially complicated arrangement of molecules and organic material, which could contribute to a greater understanding of the development and function of proteinoids. The scans also revealed anomalies in the structure of the proteinoids, which may shed light on the effects of environmental interactions on them. This could aid in the development of new proteinoids for use in medical and industrial applications by enhancing our understanding of how proteinoids react to diverse environmental conditions.

\subsection{Proteinoid L-Phe:L-Lys, L-Glu:L-Phe, L-Phe, and L-Glu:L-Phe:L-His  Electrical Oscillations in Response to Cold Light Pulses (30 minutes ON, 30 minutes OFF).}

\begin{figure}[!tbp]
\centering
\includegraphics[width=1\textwidth]{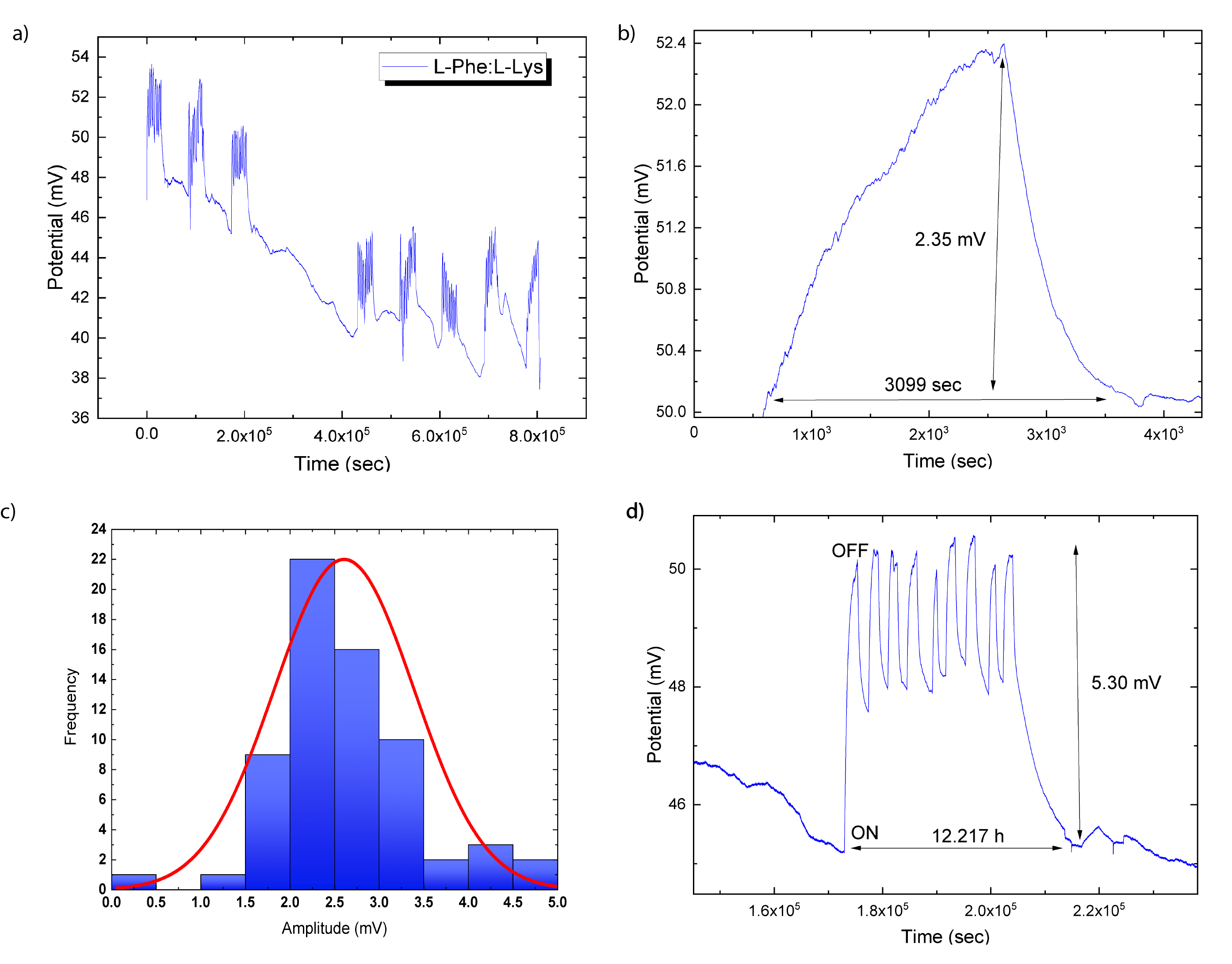}
\caption{The proteinoid's L-Phe:L-Lys electrical oscillations when exposed to cold light for 30 minutes ON and 30 minutes OFF. a) This graphic depicts a light-induced oscillation with an amplitude of 2.35 mV and a period of 3099 sec. c) Amplitudes peaked between 2-2.5 $mV$, showing a clear increase in activity. d) This illustration depicts a large oscillation amplitude of 5.30 mV and period of 12.217 hours.}
\label{ppsoiunpsssazx999}
\end{figure}

Interesting findings emerged from the study of the effects of cold white light on the proteinoid's L-Phe:L-Lys oscillations. On a 10-day timeframe, the proteinoid L-Phe:L-Lys displayed 8 large oscillations with a period of 12.2 days and an amplitude of 5.30 mV. Eight smaller oscillations, each with an amplitude of 2.35 mV and a period of 3099 sec (nealry one hour), were also detected alongside the larger ones (see Fig.~\ref{ppsoiunpsssazx999}). These findings were important because they elucidated how the proteinoid L-Phe:L-Lys reacts to room temperature in the form of visible light. Even though it had been established that proteinoids, when exposed to light, undergo oscillations, the fact that they undergo two separate oscillations with different times and amplitudes had not been previously appreciated. This finding has broad ramifications because it may shed light on the molecular mechanisms underlying proteinoid L-Phe:L-Lys. The proteinoid L-Phe:L-Lys generates smaller oscillations at a considerably faster rate than the larger ones. If the tiny oscillations are linked to the metabolic activities that take place in the proteinoid L-Phe:L-Lys, then we might be able to understand their lightning-fast tempo. Perhaps useful information could be gained regarding the biochemical pathways involved in the proteinoid L-Phe:L-Lys response to room-temperature illumination.
%
%

The summary statistics for the histogram of amplitude reveal that the average amplitude is 2.6~mV, the standard deviation is 0.78, the standard error of the mean is 0.1, the upper 95\% confidence interval is 2.8, and the lower 95\% confidence interval is 2.4. As a whole, there were 66 data points in this batch of data.
It can be seen that the majority of the data points were rather near to the mean amplitude of 2.60~mV. This indicates that the values in the dataset were spread out uniformly, with few extreme outliers. Since most of the data points were within one standard deviation of the mean and the standard deviation of the data set was 0.78, it was concluded that the data points were distributed fairly equally around the mean. With a standard deviation of 0.069 and a mean of 0.78, the fitted normal distribution places the value of $\mu$ at 2.6.

A measure of how well the fitted distribution matches the data is the -2$\times$LogLikelihood. The method relies on a comparison of the actual frequencies observed with the probabilities predicted by the fitted distribution. When this number is low, a good fit is indicated.

The Akaike Information Criterion adjusted for low sample sizes is abbreviated as AICc. It's another way to evaluate how closely a model fits the data, but it also punishes complex models with lots of parameters. To the extent that this number may be reduced, that is ideal.  Bayesian Information Criterion (BIC) is very similar to AICc, except it puts more of a damper on overly complicated models. That number ought to be as small as possible.

AICc was 158.4, while BIC was 162.6; the -2$\times$LogLikehood was 154.2. This demonstrates the validity of the normal distribution in making predictions about the amplitude data points. The $\mu$ and dispersion parameters' small errors further corroborate the normal distribution's reliability. The findings prove that the normal distribution is a good fit for the data since it permits precise estimations of the parameters governing the data central tendency and dispersion.

\begin{figure}[!tbp]
\centering
\includegraphics[width=1\textwidth]{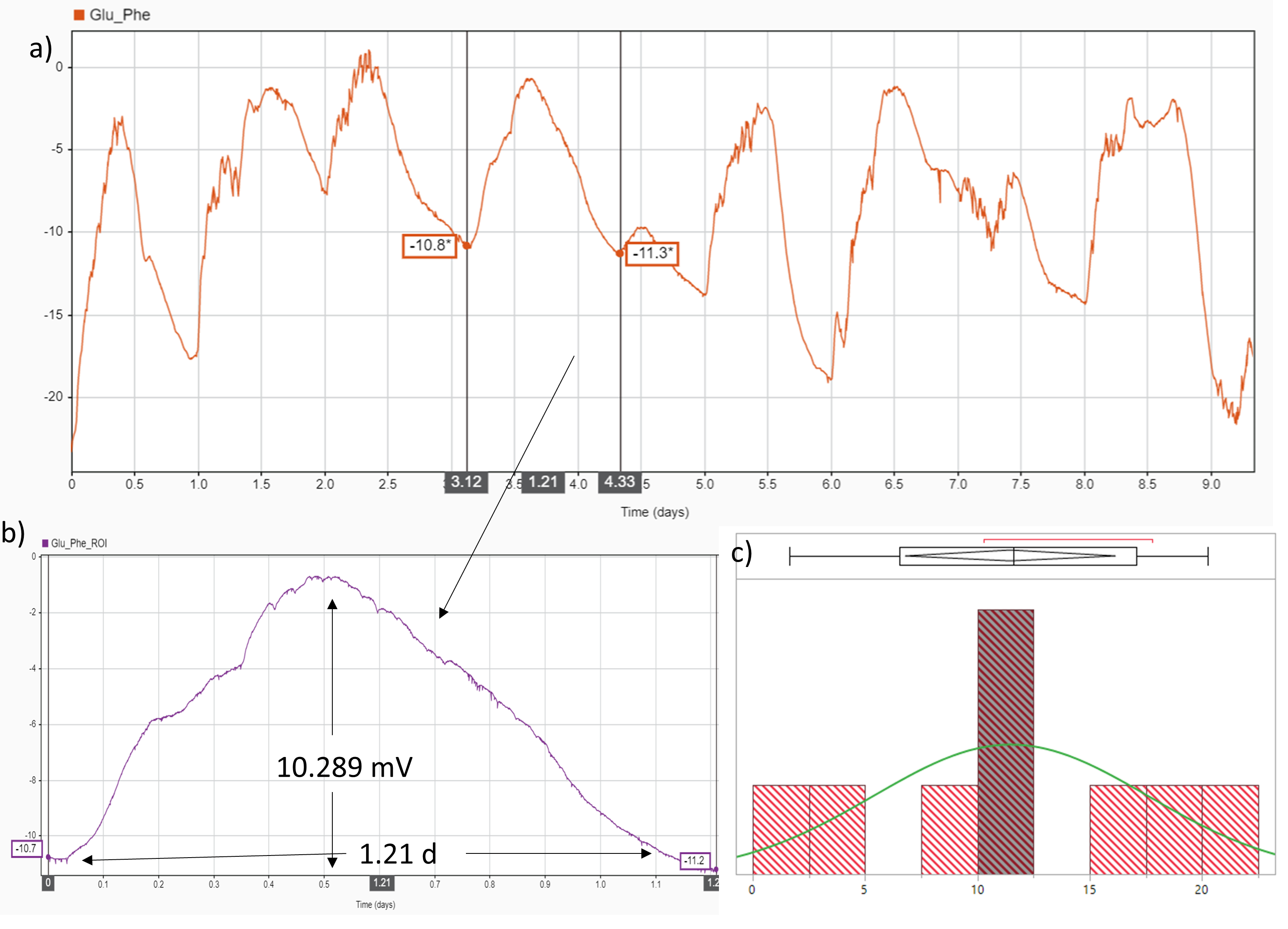}
\caption{a) The electrical oscillations of L-Glu:L-Phe proteinoid for a 10 day period, with the cold light being on for 30 min and off for 30 minutes. b) This enlarged figure shows the amplitude of 10.289 mV recorded over 1.21 days. c) This histogram and normal distribution shows the distribution of the amplitude data.}
\label{vkhgj,}
\end{figure}

White cold light has a significant impact on the amplitude and period of L-Glu:L-Phe electrical oscillations, as shown by the results of the experiment measuring this effect. The obtained data (Fig.~\ref{vkhgj,}) indicate that the average amplitude of the electrical oscillations is 11.5~mV, with a standard deviation of 6.1~mV, a standard error mean of 2~mV, and a 95\% confidence interval of 
6.8--16.2~mV. White cold light appears to have a significant impact on the oscillations, since the data show a wide range in amplitudes.

The mean, $\mu$, was estimated to be 11.4 with a standard deviation of 2. This was obtained via a normal distribution fit. With a standard error of 1.56, the dispersion was found to be 6.094. The AICc was 63.1, the BIC was 61.5, and the -2$\times$LogLikelihood was 57.1. The mean and standard deviation show that the oscillations are shifting in a more structured way than before. This is because the proteinoid is being influenced by the white cold light to oscillate in a more regular form.

\begin{figure}[!tbp]
\centering
\includegraphics[width=1\textwidth]{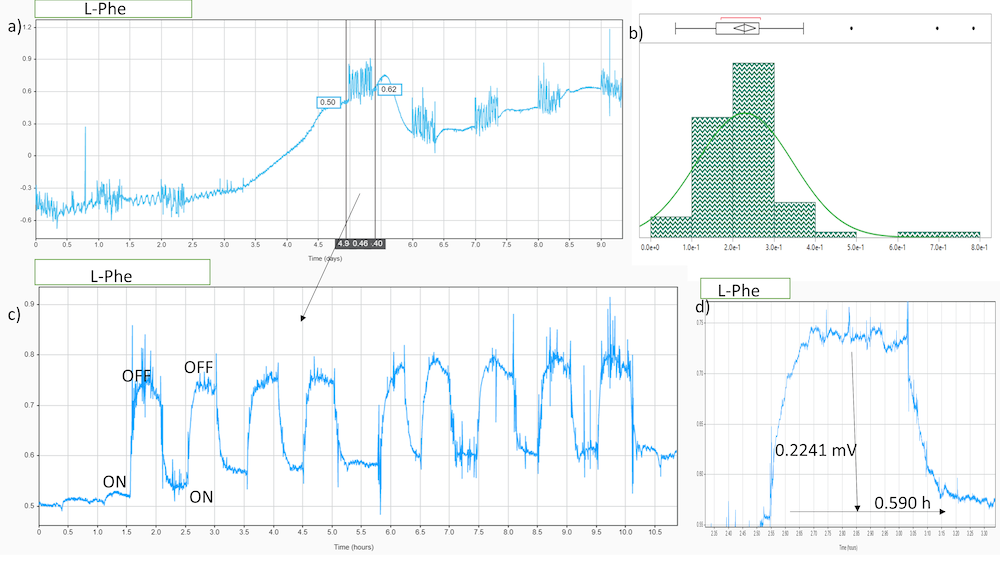}
\caption{The proteinoid electrical L-Phe oscillations after being exposed to cold light for 30~min on and 30~min off are depicted in a) this figure. The data were fit to a normal distribution, which leads us to b). c) Expanded electrical oscillations of the protenoid L-Phe. d) A spike that lasts for 35~min hours and has an amplitude of 0.2~mV.}
\label{nbdsjgva;}
\end{figure}

The average amplitude of L-Phe proteinoid's electrical oscillations has been measured (Fig.~\ref{nbdsjgva;}). The research found that the average amplitude of L-Phe proteinoids was 0.2~mV, with a standard deviation of 0.1~$mV$ and a mean standard error of 0.01~mV. The 95\% confidence interval ranged from a high of 0.26 to a low of 0.20. A total of 73 data points were included in the study. The amplitude and dispersion of the normal distribution were calculated to be $\mu$=0.23 and $\sigma$=0.13, respectively. In this case, the -2$\times$LogLikelihood=-107.63, AICc=-103.5, and BIC=-99.1 are all negative numbers.

\begin{figure}[!tbp]
\centering
\includegraphics[width=1\textwidth]{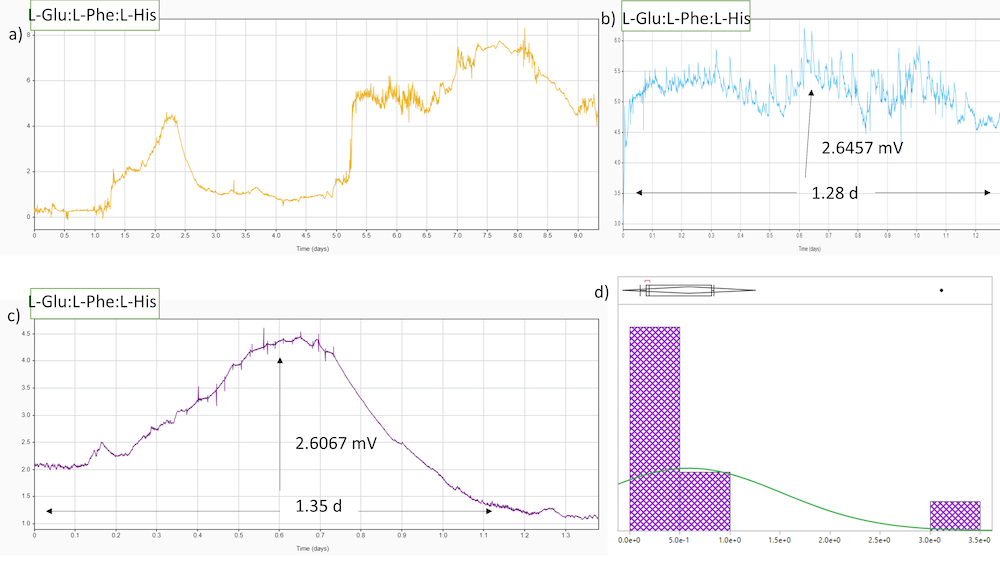}
\caption{a) The electrical oscillations of L-Glu:L-Phe:L-His grew and shrank in step with the periodic exposure to cold light. b) With time, the volume and intensity of seemingly random, tiny electrical oscillations increased. c) An electrical oscillation of 2.6~mV amplitude and 1.35 days period was found. By d) fitting the data to a normal distribution, an accurate picture of the fluctuations was created.}
\label{jdsfhbjkghas;hg;}
\end{figure}

In reaction to light, the L-Glu:L-Phe:L-His proteinoid does not appear to oscillate in a regular fashion, as shown by the experimental data (Fig.~\ref{jdsfhbjkghas;hg;}). The oscillations averaged 0.6~mV in amplitude, with a standard deviation of 0.9~mV and a standard error of the mean of 0.3~mV. There was a 95\% confidence interval ranging from 1.3~mV to -0.007~mV. The standard deviation of the fitted normal distribution was 0.3, whereas the dispersion was 0.93 (with an error of 0.23). 25.9 was the value for -2$\times$logLikelihood, 31.6 was the value for the AICc, and 30.6 was the value for the BIC. The findings imply that the proteinoid L-Glu:L-Phe:L-His does not behave predictably when exposed to different intensities of white light. Fitted normal distribution showed modest mean amplitude and narrow dispersion. The proteinoid's structure complexity may be to blame for this.

\subsection{Impact of Continuous White Cold Light Exposure on the Electrical Oscillations of Proteinoids L-Phe:L-Lys,L-Glu:L-Phe,L-Phe, and L-Glu:L-Phe:L-His.}

\begin{figure}[!tbp]
\centering
\includegraphics[width=1\textwidth]{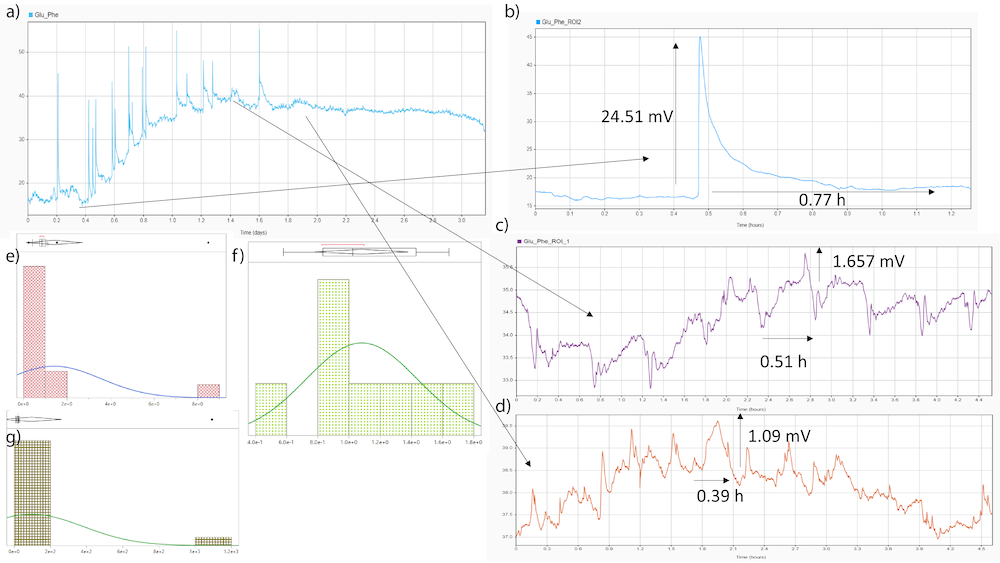}
\caption{(a) The electrical oscillations of the proteinoid L-Glu:L-Phe when it is constantly illuminated by cold white light; (b,c,d) magnifications of different regions of electrical oscillation according to (a); and (e,f,g) a histogram of the three different regions of spikes produced by L-Glu:L-Phe. }
\label{vgffjjhlj;hg;}
\end{figure}

Experiments with the proteinoid L-Glu:L-Phe under constant exposure to cold light reveal three different types of spikes (Fig.~\ref{vgffjjhlj;hg;}). The large spikes are followed by smaller ones, and the average amplitude of the first sequence (Fig.~\ref{vgffjjhlj;hg;}b) of spikes is 95.8~mV, with a 95\% confidence interval spanning from 261.8~mV to -70.1~mV. The $\mu$=95.8 centre and $\sigma$=287.3 dispersion are those of the fitted normal distribution. There is a -2$\times$LogLikelihood, AICc, and BIC of 197.2, 202.3, and 202.5, respectively.
According to the proteinoid L-Glu:L-Phe, the second region's spikes (Fig.~\ref{vgffjjhlj;hg;}c) have a mean value of potential 1.42 mV, with a 95\% higher mean of 0.59 and a 95\% lower mean of 0.12. The parameters obtained from the fitted normal distribution are as follows: mean = 1.42, standard deviation = 2.15, location $\mu$=1.42, dispersion $\sigma$=2.15, -2$\times$LogLikelihood = 54.8, AICc = 60.995, and the BIC = 60.93. These findings suggest that the second region's spikes follow a normal distribution, with a mean and variance that are typical of this type of distribution.
Spikes in the third area (Fig.~\ref{vgffjjhlj;hg;}d) have a mean potential of 1.08 mV, with a 95\% upper limit of 1.38 mV and a 95\% lower limit of 0.783 mV, as measured by the proteinoid L-Glu:L-Phe. According to the normal distribution fit, the following values were found to be appropriate: mean = 1.08, standard deviation = 0.36, location $\mu$ = 1.08$\pm$0.13, dispersion $\sigma$ = 0.36$\pm$0.01, -2$\times$LogLikelihood = 5.26, AICc = 11.7, and the BIC = 9.42. According to these results, the second region's spikes have the mean and variance expected of a normal distribution.

\begin{figure}[!tbp]
\centering
\includegraphics[width=1\textwidth]{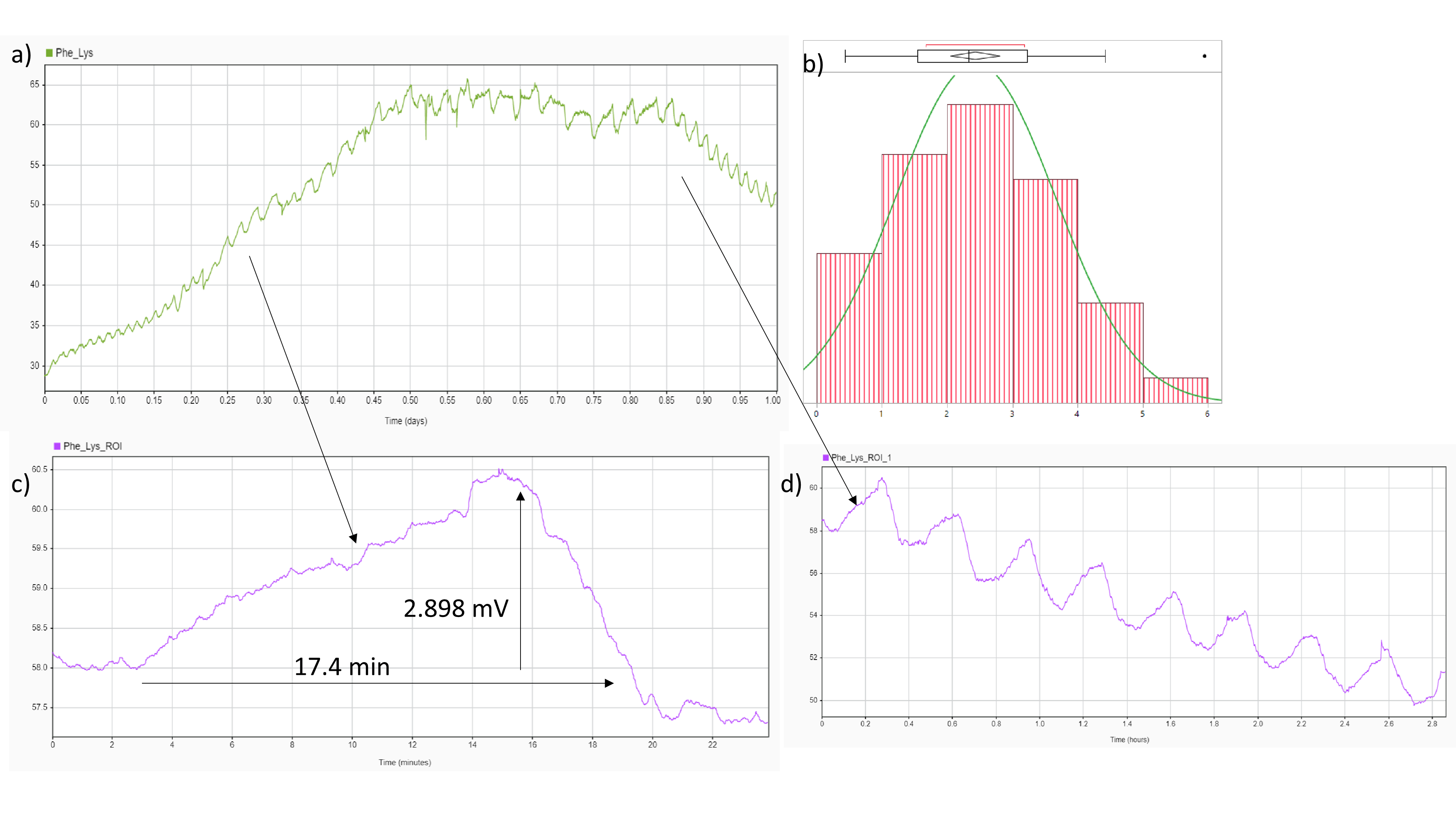}
\caption{a) plots the potential oscillations in mv against days of proteinoid L-Phe:L-Lys 
b) An electrical oscillation with an amplitude of 
2.9~mV and a period of 17.4 minutes is magnified and plotted against time in the figure. 
c) The amplitude of the potential was suited by a normal distribution with the centre at 2.4$\pm$0.2 and the dispersion $\sigma$ at 1.2$\pm$0.14 
d) Example of a typical oscillation pattern.
 }
\label{vgffjadsgsdgs;}
\end{figure}

Figure~\ref{vgffjadsgsdgs;} shows that the proteinoid L-Phe:L-Lys, when exposed to cold white light over the course of the tests, consistently displays smaller, repeatable patterns. Standard deviation is 1.23 and standard error of the mean is 0.19 for an amplitude of 2.4~mV. The upper 95\% confidence interval for amplitude was 2.82 $mV$, and the lower 95\% confidence interval was 2.05 $mV$. Location $\mu$= 2.44$\pm$0.19 and dispersion $\sigma$=1.227$\pm$0.136 are estimated using the fitted normal distribution. The calculated -LogLikelihood=153.36 was accompanied by AICc=139.7 and BIC=142.8. 

\begin{figure}[!tbp]
\centering
\includegraphics[width=1\textwidth]{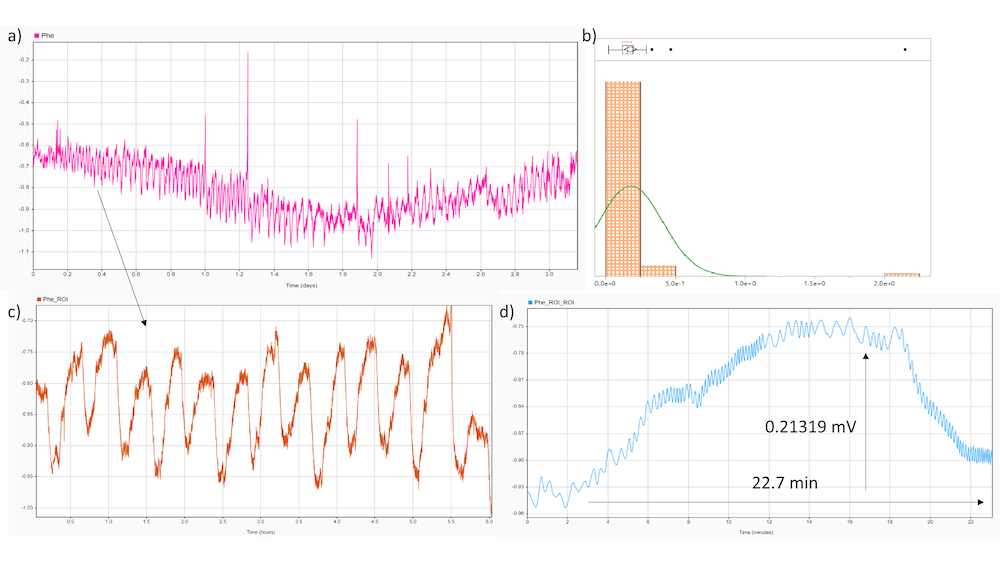}
\caption{a) Potential oscillations of an L-Phe proteinoid are shown to be occurring under the constant impact of cold light, and b) the amplitude distribution is shown to be right skewed, with a dispersion of $\sigma$=0.23$\pm$0.02 and a location of $\mu$=0.18$\pm$0.03.c) L-Phe spiking activity pattern. An example d) spike with an amplitude of 0.2~mV and a period of 22.7 min.
 }
\label{vgmkdsnvknjvn;sak}
\end{figure}

Constant exposure to cold white light significantly affected the electrical oscillations of the potential of the L-Phe protenoid (Fig.~\ref{vgmkdsnvknjvn;sak}). The standard deviation of the amplitude was 0.18 millivolts. Mean 0.025, 95\% confidence interval (CI) 0.23 to 0.13, N=83 observations. Position $\mu$=0.18$\pm$0.03 and dispersion $\sigma$=0.23$\pm$0.02 are calculated from the normal distribution that best fits the data. -2$\times$LogLikehood=-10.51, AICc=-6.36, and BIC=-1.67 are all quantifications of the normal distribution.

\begin{figure}[!tbp]
\centering
\includegraphics[width=1\textwidth]{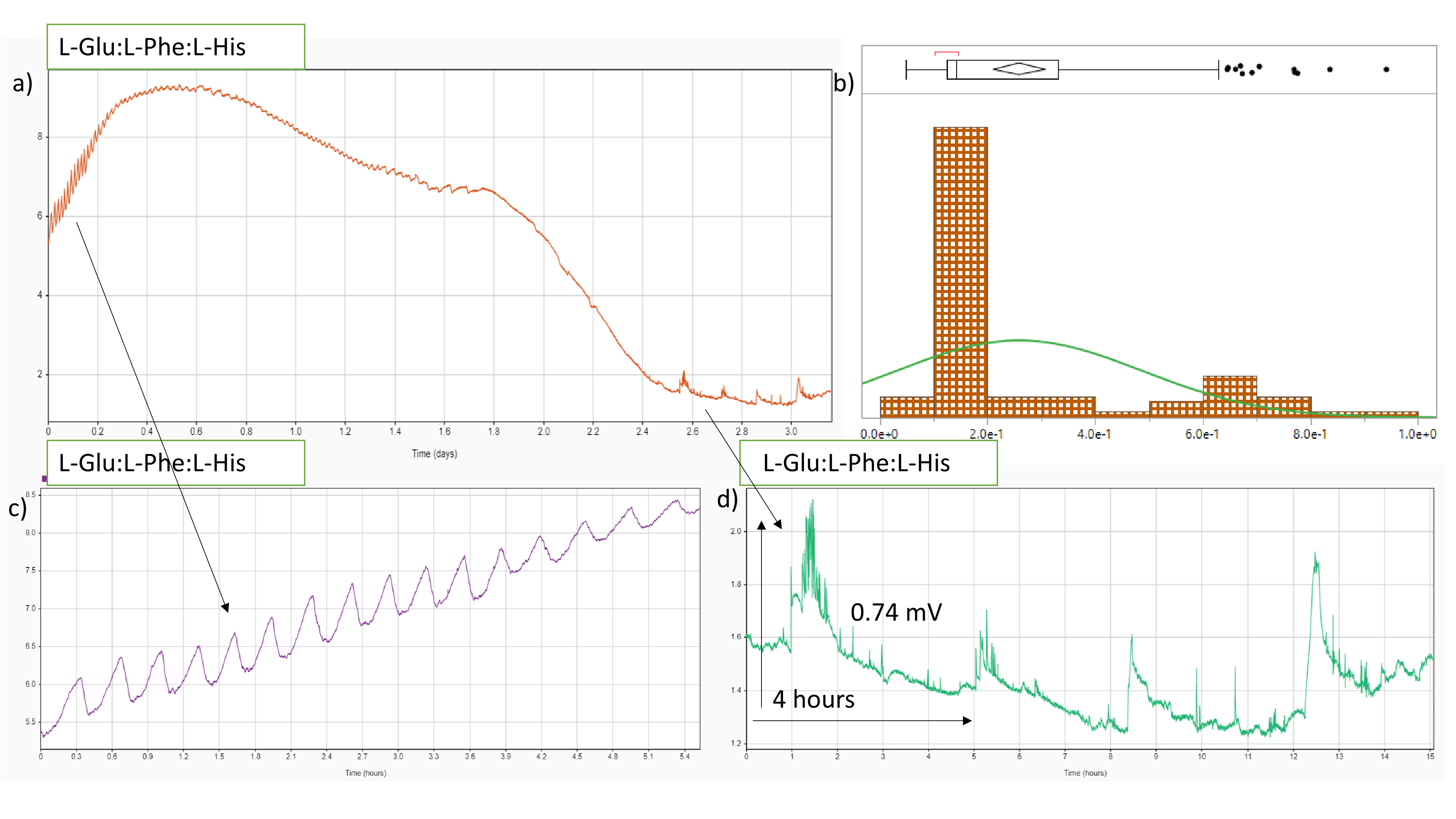}
\caption{ As seen in a), the proteinoid L-Glu:L-Phe:L-His exhibited electrical oscillations when continuously exposed to cold light. b) Location $\mu$ = 0.26$\pm$0.02, Dispersion $\sigma$=0.23$\pm$0.02, and a Normal Distribution Curve Fitted to the Amplitude of the Potential Histogram. Amplification of Potential, in mv (c) Unique spikes with an amplitude of 0.74~mV and a duration of 4 hours are shown in (d).
 }
\label{fsdgsaghahafhafhdafha;}
\end{figure}

The average amplitude of L-Glu:L-Phe:L-His (Fig.~\ref{fsdgsaghahafhafhdafha;}) was 0.3~mV, with a standard deviation of 0.02~mV. With a sample size of 87, the 95th percentile mean fell between 0.31 and 0.21. The location $\mu$ was 0.29 with a standard error of 0.02, and the dispersion $\sigma$ was 0.23 with a standard error of 0.018, both according to the fitted normal distribution. The results were a BIC value of -2.12, an AICc value of 6.91, and a -2$\times$LogLikelihood value of -11.1. All of the fit metrics show that the distribution is a good one. This indicates that L-Glu:L-Phe:L-His is an amplitude-consistent and robust proteinoid.

\subsection{Examining Variations in Proteinoid Electrical Oscillations with Cold Light}

\begin{figure}[!tbp]
\centering
\includegraphics[width=1\textwidth]{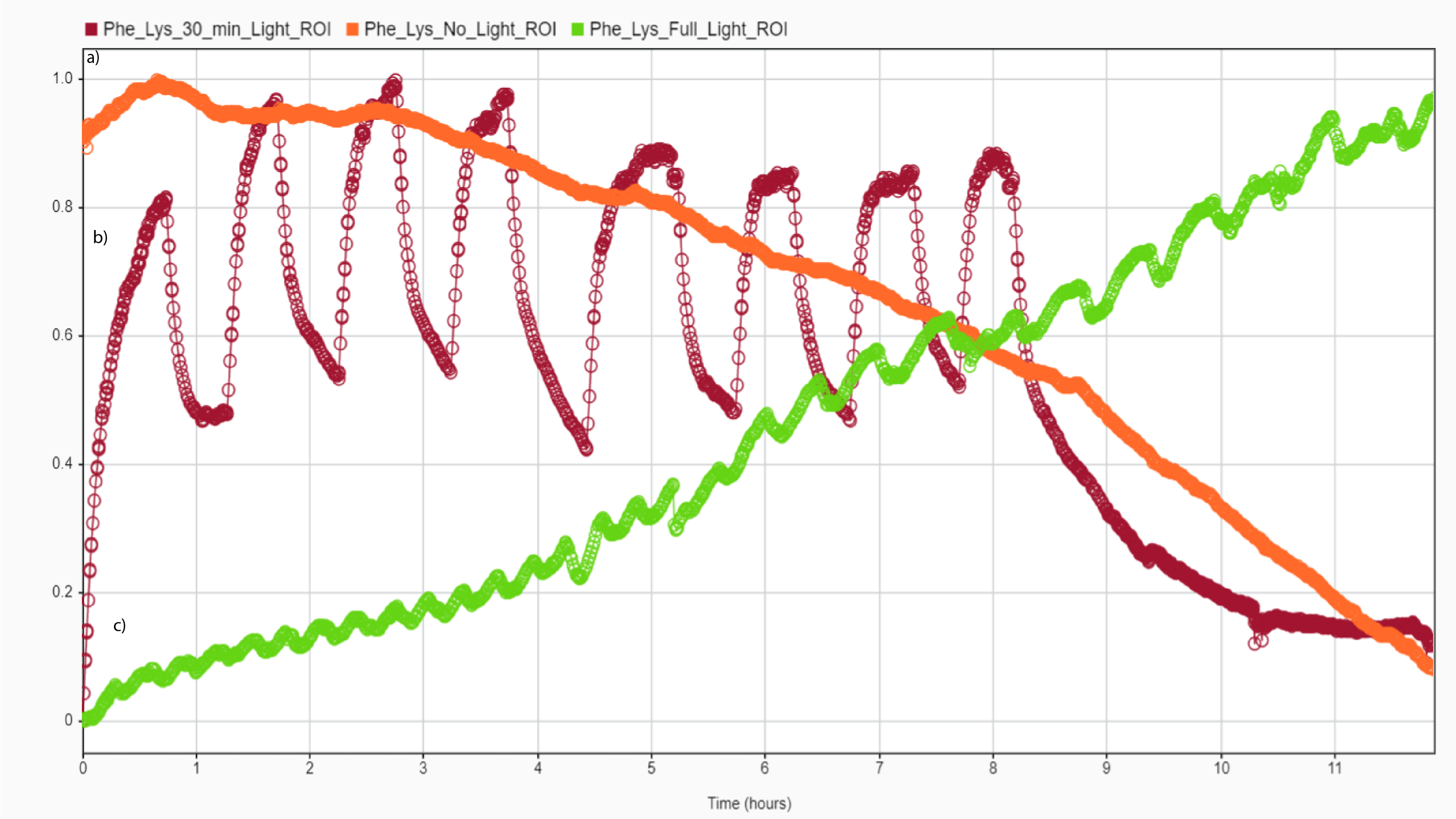}
\caption{Presented here is a schematic depicting the potential amplitude (normalised Y axis values) for the proteinoid L-Phe:L-Lys. When a) no light is present, and b) white, cold white light is on for 30 minutes, then off for 30 minutes. b) When the cold light is on during the entire time of the experiment. }
\label{kfjalkjfksakglksg;}
\end{figure}

\begin{figure}[!tbp]
\centering
\includegraphics[width=0.8\textwidth]{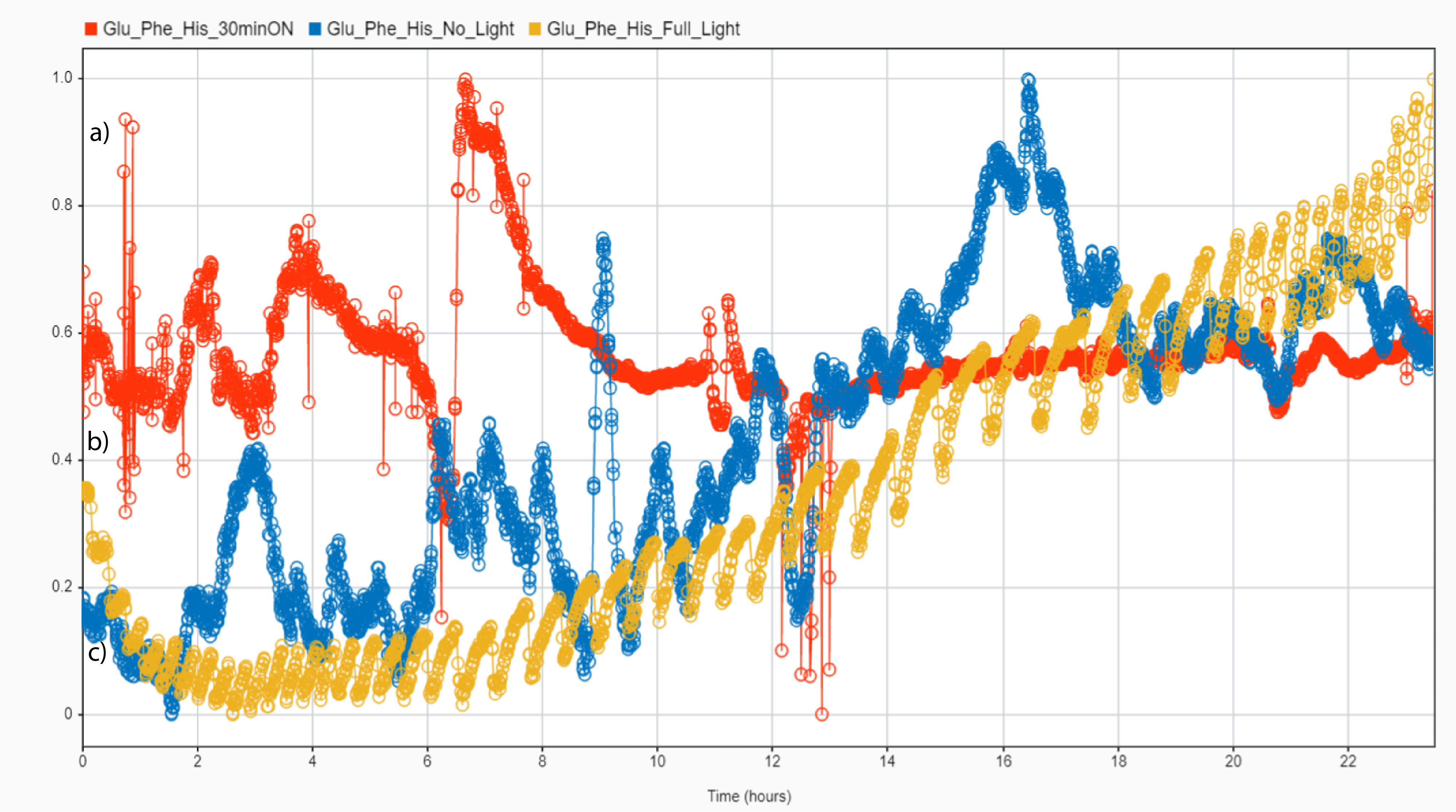}
\caption{ Electrical oscillations of L-Glu:L-Phe:L-His are depicted in the picture when a) white light is on and off for 30 minutes in a row. White light illumination is absent (b). c) When the entire experiment is conducted in cold white light. }
\label{gdshsjg;}
\end{figure}

The results of the experiment with the proteinoids L-Phe:L-Lys (Fig.~\ref{kfjalkjfksakglksg;}) and L-Glu:L-Phe:L-His (Fig.~\ref{gdshsjg;}) showed that the proteinoids show small, quick oscillations when the light is on, but large, slow peaks when the light is on-off for 30 minutes (Table~\ref{fdjdsjsgj}). In the realm of logical gates and alternative computing, this discovery has far-reaching applications. 
However, the L-Glu:L-Phe-L-His proteinoid differed from the other in that it exhibited oscillations even in the absence of white light for the first hour of the experiment. Proteinoids, which can be trained to behave in a predetermined fashion to external stimuli like light, make excellent candidates for building logic gates. The ability to process information and make decisions based on specific inputs is the basis for the creation of logical gates that exhibit this behaviour.

\begin{figure}[!tbp]
\centering
\includegraphics[width=1\textwidth]{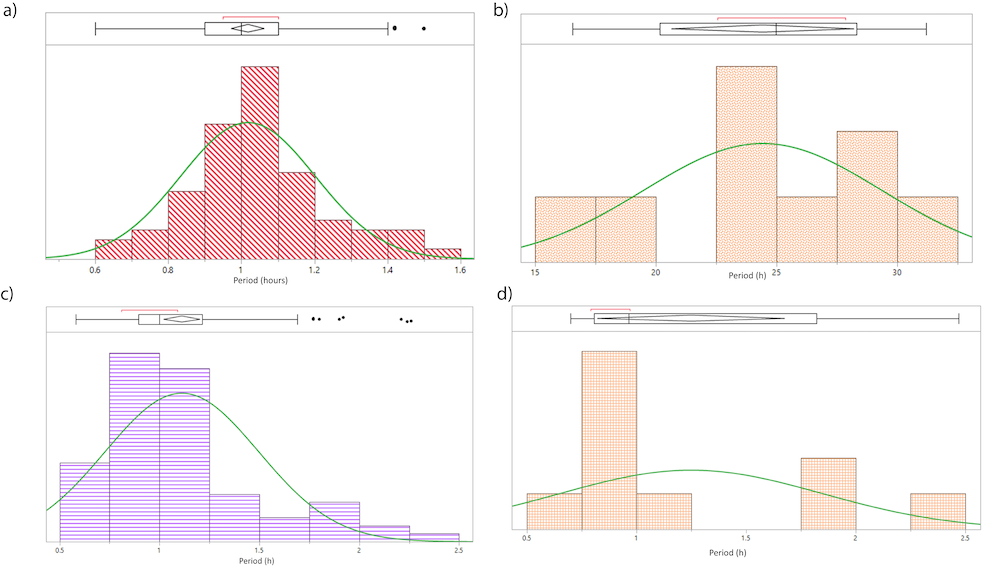}
\caption{Electrical oscillations' period histograms and fitted distributions are displayed in the figure. a) L-Phe:L-Lys proteinoid mean period is 1.02$\pm$0.18 h, location $\mu$=1.02$\pm$0.02, dispersion $\sigma$= 0.18$\pm$0.02. b) L-Glu:L-Phe proteinoid mean period is 24.4$\pm$4.9 h, location $\mu$=24.4$\pm$1.6, dispersion $\sigma$=4.9$\pm$1.2, c) L-Phe proteinoid mean period is 1.1$\pm$0.38, location $\mu$=1.11$\pm$0.045, dispersion $\sigma$=0.38$\pm$0.03. d) L-Glu:L-Phe:L-His proteinoid  mean period is 1.25$\pm$0.6, location $\mu$=1.25$\pm$0.19 and dispersion $\sigma$=0.59$\pm$0.15.  }
\label{cmdec;}
\end{figure}

\begin{table}[!tbp]
\centering
\caption{The mean amplitude, mean period data for illuminated proteinoids L-Phe:L-Lys, L-Glu:L-Phe, L-Phe, and L-Glu:L-Phe:L-His.}
\begin{tabular}{|c|c|c|c|}
	\hline
	Proteinoid  & Light & Mean & Mean      \\
       &  Condition  &  Potential  & Period      \\
    &  & (mV) & (h)      \\
	\hline\hline
	L-Phe:L-Lys & Periodic & 2.61$\pm$0.78  & 1.02$\pm$0.18       \\
	\hline
	L-Glu:L-Phe & Periodic  & 11.48$\pm$6.09 &  24.43$\pm$4.93      \\
	\hline
	L-Phe & Periodic & 0.23$\pm$0.12  & 1.1$\pm$0.38     \\
	\hline
	L-Glu:L-Phe:L-His & Periodic & 0.59$\pm$0.93 & 1.25$\pm$0.6     \\
	
	\hline
	\end{tabular}
\label{fdjdsjsgj}
\end{table}

According to the data in Fig.~\ref{cmdec;} and Tab.~\ref{fdjdsjsgj}, when exposed to periodic white light, L-Glu:L-Phe proteinoids display oscillation durations 24 times longer than those of L-Phe:L-Lys, L-Phe, L-Glu:L-Phe:L-His. This is because L-Glu:L-Phe proteinoids have a higher concentration of the essential amino acids glutamic acid and phenylalanine. While phenylalanine serves as a hydrophobic core when combined with other amino acids, glutamic acid is well-known as a potent chelating agent~\cite{shi2020effects} that binds strongly to metal ions. Higher oscillation periods are indicative of a greater affinity for metal ions in the L-Glu:L-Phe proteinoids, which in turn allows them to absorb more light energy. The benefits of this longer time of oscillations are numerous. It improves the efficiency with which the L-Glu:L-Phe proteinoid can take in energy, which in turn may increase the number of chemical reactions taking place within the molecule. This can result in more rapid and effective utilisation of energy, which in turn can improve metabolic processes. Because the proteinoid can react more quickly to changes in its environment, a longer oscillation period can also make it more sensitive to such changes. The significance of this finding rests in the fact that the L-Glu:L-Phe proteinoid is more sensitive to periodic light than the other three proteinoids, suggesting that it may be more suited for use in light-activated switches, sensors, and other light-sensitive devices.

The peak-to-peak (pk-pk) ratio of a waveform is defined as the ratio of its maximum positive amplitude to its minimum negative amplitude. Without a DC component, the amplitude of an AC wave at its positive and negative peaks (peak tp peak) is twice that of the positive peak.
When measuring electrical oscillations, the difference between the highest and lowest points is known as peak-to-peak voltage ($V_{p-p}$). Multiplying the RMS voltage ($V_{rms}$) by 2$\times$$\sqrt{2}$ yields the ($V_{p-p}$) voltage (Eq.~\ref{eq2aaa02}).
\begin{equation} \label{eq2aaa02}
V_{p-p}= 2\sqrt{2} \times V_{rms}
\end{equation}
The RMS voltage represents the effective value of an oscillation. It can be derived from the peak voltage ($V_{p}$) by dividing it by $\sqrt{2}$ (Eq.~\ref{eq2aaa02})~\cite{RMS}.
\begin{equation} \label{eq2aaa02}
V_{p}=  \frac{V_{rms}}{\sqrt{2}}
\end{equation}

\begin{figure}[!tbp]
\centering
\includegraphics[width=1\textwidth]{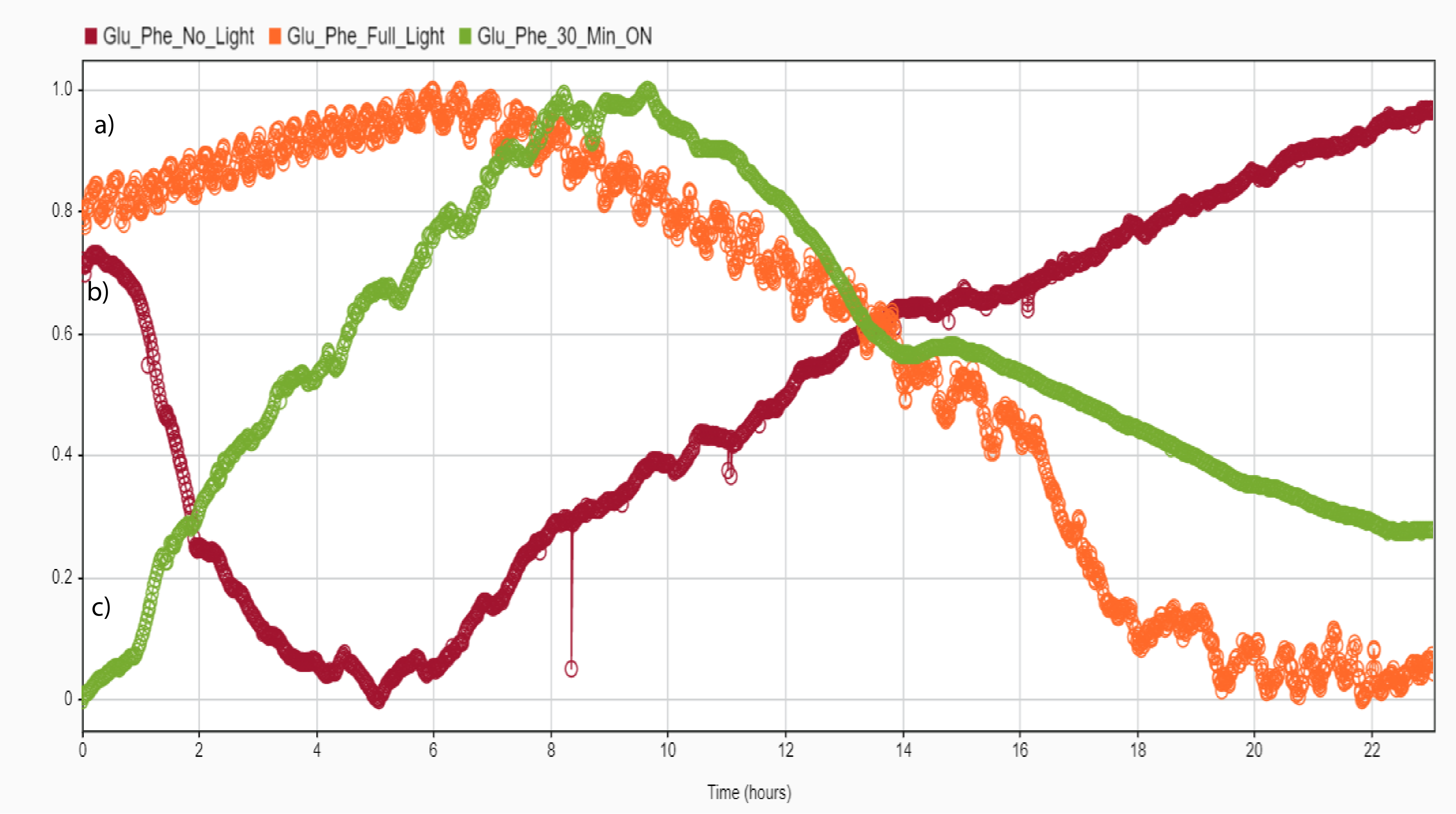}
\caption{The electrical oscillations of the proteinoid L-Glu:L-Phe a) in the presence of white light during the experiment. b) When there is no illumination. c) When white light is on for 30 minutes and then off for 30 minutes.
   }
\label{gdshssdsabjg;}
\end{figure}

Experiments measuring the peak-to-peak ratio of L-Glu:L-Phe oscillations under cold light, 30 minutes on, 30 minutes off, and full illumination yielded values of 2.67, 1.25, and 3.80E-3, respectively, after being run for one hour (Fig.~\ref{gdshssdsabjg;}). Useful for a variety of scenarios, this could inform future research into the L-Glu:L-Phe proteinoid and how it behaves in various settings. This study has dual implications. It's useful since it explains how the L-Glu:L-Phe proteinoid works and how it can be tweaked to get the effects we want. It may be possible to optimise the proteinoid's performance in biomimetic computing by modifying its oscillations if one has a thorough understanding of the effects of light and the ratio between peak to peak. This work contributes to our theoretical understanding of light's function at the cellular and molecular levels. Furthermore, this finding may find use in biotechnology. If we want to increase the efficiency of a certain biochemical process or change the behaviour of a certain protein, we could use white light to modify the oscillations of L-Glu:L-Phe proteinoids, for example.

\begin{figure}[!tbp]
\centering
\includegraphics[width=1\textwidth]{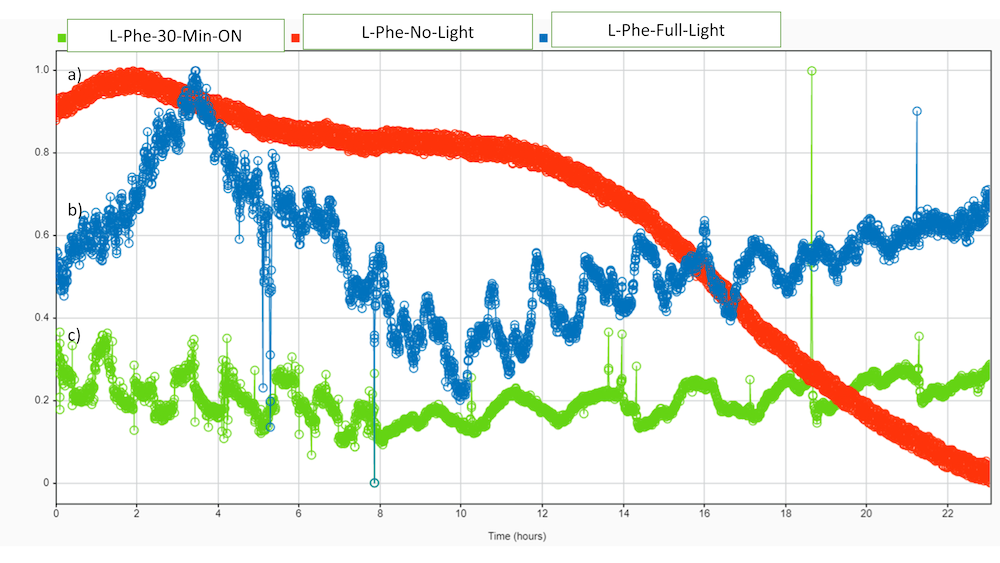}
\caption{Proteinoid L-Phe's electrical oscillations.
a) During 30 minutes of exposure to cold light. b) When room temperature light was used to illuminate the sample. When (c) a steady, white, cold light is illuminating on the sample. }
\label{jkfbdsadg;asdj;}
\end{figure}

Furthermore, this research shows that L-Phe electrical oscillations (Fig.~\ref{jkfbdsadg;asdj;})  behave differently under constant and varied lighting conditions. The proteinoid L-Phe, when illuminated continuously, displayed a periodic oscillation pattern of relatively small amplitude and period. When the light was cycled on and off every thirty minutes, however, the oscillations increased in amplitude and time. When the cold light was turned off, no oscillations were seen (Fig.~\ref{jkfbdsadg;asdj;}).

The peak-to-peak ratio of L-Phe oscillations exposed to cold light continuously versus oscillations exposed to white light for 30 minutes on and 30 minutes off was determined to be 1.45/1000 by this experiment. Differences in the amplitude of the oscillations between the two states are so evident. The results of this study have significant bearing on the future of computing outside the norm. Modulating and controlling the level of oscillations through exposure to white light could lead to the development of innovative computer systems. Additionally, this degree of modulation is possible with a comparatively minimal amount of energy, which suggests it may contribute to a decrease in computing's energy costs. As a conclusion, this modulation may pave the way for the creation of more complex computing applications, such as those that necessitate fine-grained control of oscillations.

These results indicate that electrical oscillations based on proteinoids may be particularly sensitive to variations in light intensity. Perhaps this arises from the proteinoid's photoreactive characteristics, which are altered in response to illumination. Additionally, the results of this research show that proteinoids may provide a useful foundation upon which to build new kinds of light-sensitive electrical oscillators. 

\subsection{Regulation of Proteinoid Oscillation Amplitudes by Light Intensity
}

\begin{figure}[!tbp]
\centering
\includegraphics[width=1\textwidth]{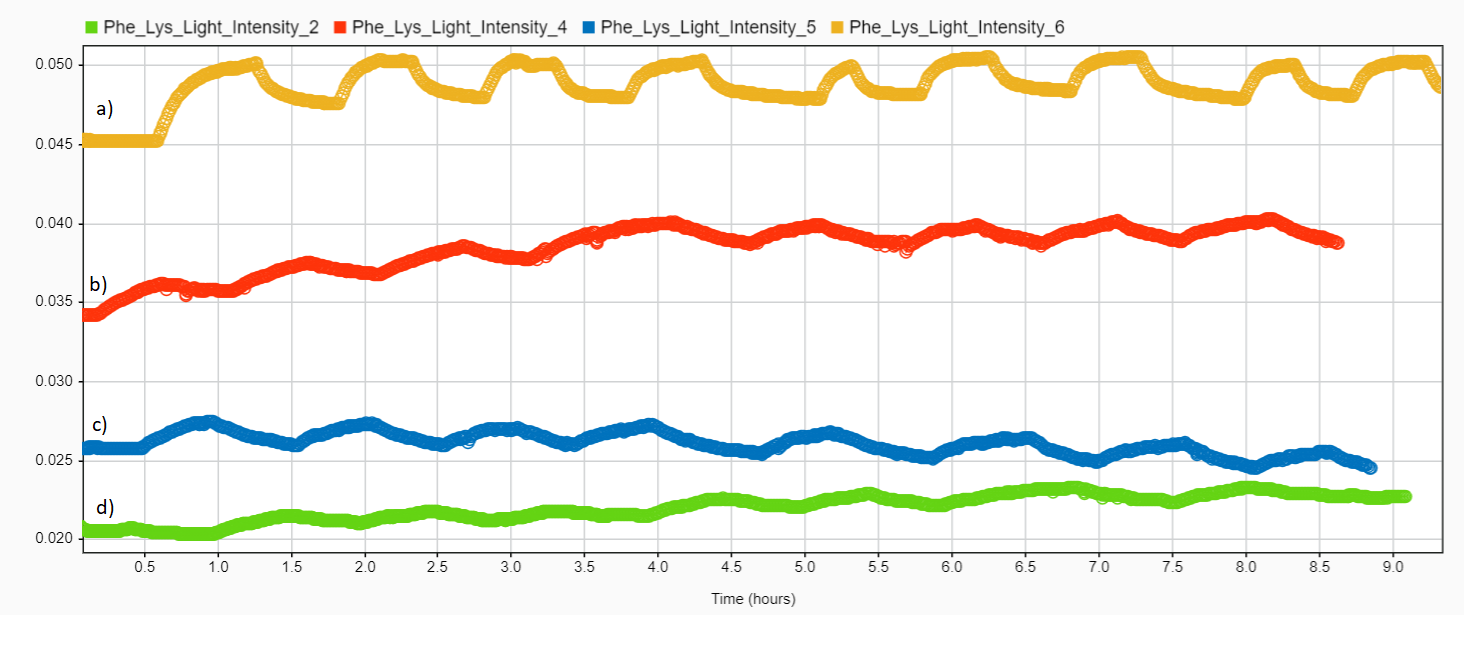}
\caption{The variation in potential amplitude for L-Phe:L-Lys proteinoid as a function of illumination level. Intensity levels range from a) 186.6~klux to c) 134.2~klux to d) 55.9~klux.
}
\label{sadasd;}
\end{figure}

\begin{figure}[!tbp]
\centering
\includegraphics[width=1\textwidth]{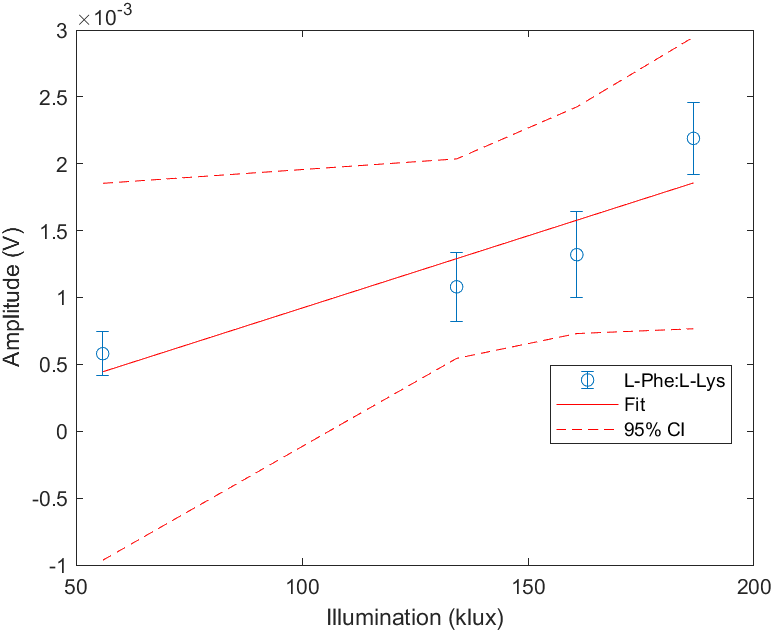}
\caption{The linear fit of the L-Phe:L-Lys proteinoid amplitude variance while illuminated with light intensities of 68~klux, 100.5~klux, 132.8~klux, and 180.3~klux demonstrates, with a 95\% confidence interval, the proteinoids' strong and repeatable sensitivity to light intensity.
}
\label{jkfbdsa;}
\end{figure}

\begin{table}[!tbp]
\centering
\caption{With an estimated slope of x1=1.0804e-5, SE of 3.5407e-6, tStat of 3.0515, and pValue of 0.0927, the linear fit model of proteinoid L-Phe:L-Lys demonstrates a significant relationship between light intensity and potential oscillation amplitude.}
\label{light_amplitude_approximation}
\begin{tabular}{|c|c|c|c|c|}
	\hline
	  & Estimate & SE & tStat & pValue     \\
      &    &    &   &    \\
   	\hline\hline
	(Intercept) & -1.5933e-4 & 5.0636e-4  & -0.3146&  0.7828        \\
	\hline
	x1 & 1.0804e-5  &3.5407e-6 &  3.0515 & 0.0927     \\
		
	\hline
	\end{tabular}
\label{fdjdsjsgj}
\end{table}

When irradiated with varying light intensity, the oscillations of the proteinoids L-Phe:L-Lys depend on the type and wavelength of light employed. When proteinoids are irradiated with white light, their potential fluctuates depending to the light's intensity, with greater intensities resulting in pronounced oscillations. For instance, when proteinoids L-Phe:L-Lys are illuminated with 68~klux, the potential has a significantly smaller amplitude comparing to that of proteinoids exposed to 100.5~klux (Figs.~\ref{sadasd;} and \ref{jkfbdsa;}). An approximate of a relationship between light intensity and proteinoids' potential oscillation amplitude is presented in Tab.~\ref{fdjdsjsgj}.

\section{Discussion}

When irradiated with cold white light, proteinoids produce oscillations of the potential measured between two elecrode insert in the solution with proteinoids. If the cold light is always on, the amplitude of oscillations decreases. This indicates that the proteinoids are responsive to variations in lighting conditions and that the oscillations are set off by the on/off nature of the light source. We still don't know how this phenomenon occurs. The proteinoids may have an internal oscillator that is set off by the repetitive on and off of the light source. As a timer, this oscillator allows the proteinoids to react predictably to changes in illumination. The oscillator is not activated while the light is on continuously, therefore the waves get weaker and more frequent. Proteinoids may instead be reacting to the level of illumination rather than the on/off pattern. When the light is on continuously, with constant intensity, the amplitude of the oscillations decreases but the frequency increases. This indicates the proteinoids are capable of responding to varying levels of illumination. It's also possible that the proteinoids respond not just to the brightness of the light but also to its temporal pattern. Proteinoids may be able to detect a pattern when the light is flashed on and off. When the light is on all the time, it throws off the temporal pattern, causing the amplitude and frequency of the oscillations to change. 

\begin{figure}[!tbp]
\centering
\includegraphics[width=1\textwidth]{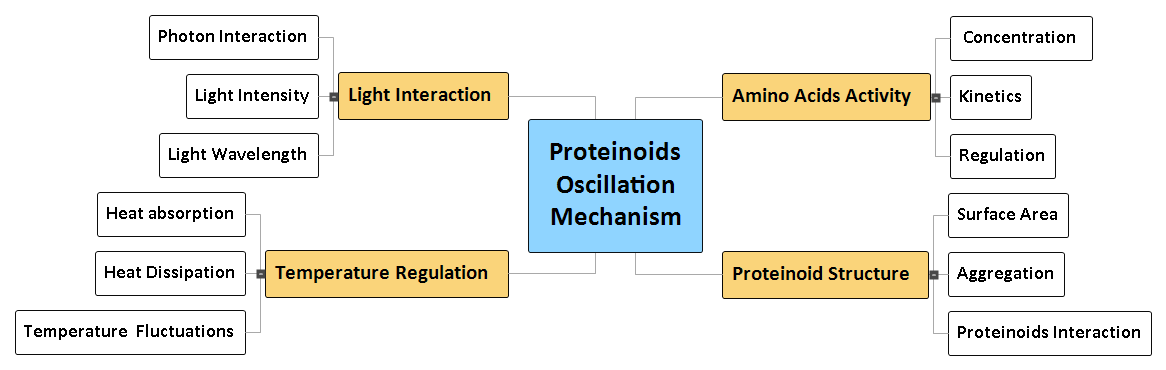}
\caption{ Light-induced oscillations in the proteinoids provide evidence that these molecules may serve as light-sensitive switches.}
\label{jkfbdsaasdda;}
\end{figure}

The possible mechanisms that could account for the observed oscillations are laid forth in the preceding mind-map (Fig.~\ref{jkfbdsaasdda;}). A number of factors, such as the function of individual amino acids, the structure of proteinoids, the effects of heat and light, and so on, may be at play. These phenomena may have significant ramifications for computational systems like quantum and biological computers; therefore, more study is needed to identify the specific mechanisms responsible for the observed oscillations. Proteinoids are a tool for manipulating cellular behaviour in biological computing. Proteinoid oscillations can be used to coax cells into making desired substances like hormones and enzymes~\cite{kolitz2018recent}. This has applications in the development of AI and handling of robotic systems.

Proteinoid oscillations triggered by cold light are an intriguing phenomenon with potential uses in both human cerebral activity~\cite{messe2014relating,haeger2019cerebral} and photosynthesis~\cite{gentili2020light}. Proteinoids exposed to cold light are thought to generate electrical potential oscillations that could be utilised to mimic neuronal activity in the brain. This may help in the study of mental illnesses including depression and dementia. 

\section{Conclusions}

In conclusion, our study shows that white cold light affects proteinoids' amplitude, oscillating pattern, and spiking behaviour. Potential uses in unconventional computing using proteinoid spikes as optical sensors are now possible. There is a need for more study into the mechanism behind light-induced spiking in proteinoids and their potential applications.

\section*{Acknowledgement}

The research was supported by EPSRC Grant EP/W010887/1 ``Computing with proteinoids''. Authors are grateful to David Paton for helping with SEM imaging and to Neil Phillips for helping with instruments.



\end{document}